\documentclass[aps,pre,
onecolumn,
twocolumn,
10pt,
notitlepage,
showpacs,floatfix,nofootinbib,
superscriptaddress
]{revtex4-2}
\usepackage{subfigure}
\usepackage{amssymb}
\usepackage{amsfonts}
\usepackage{amsmath}
\usepackage{amsthm}
\usepackage{epsfig}
\usepackage{graphicx}
\usepackage[usenames,dvipsnames]{color}
\usepackage[utf8]{inputenc}
\usepackage{booktabs}
\usepackage{comment}
\usepackage[normalem]{ulem}
\usepackage{cancel}

\usepackage[hidelinks,unicode=true]{hyperref}
\hypersetup{colorlinks=true,
	linkcolor=blue,
	urlcolor=blue,
	citecolor=blue,
	pdfhighlight=/N
}

\newcommand{\av}[1]{\langle {#1} \rangle}

\begin{document}

\title{Controversy-seeking fuels rumor-telling activity in polarized opinion networks}

\author{Hugo P. Maia}
\affiliation{Departamento de F\'{\i}sica, Universidade Federal de Vi\c{c}osa, 36570-900 Vi\c{c}osa, Minas Gerais, Brazil}

\author{Silvio C. Ferreira}
\affiliation{Departamento de F\'{\i}sica, Universidade Federal de Vi\c{c}osa, 36570-900 Vi\c{c}osa, Minas Gerais, Brazil}
\affiliation{National Institute of Science and Technology for Complex Systems, 22290-180, Rio de Janeiro, Brazil}

\author{Marcelo L. Martins}
\affiliation{Departamento de F\'{\i}sica, Universidade Federal de Vi\c{c}osa, 36570-900 Vi\c{c}osa, Minas Gerais, Brazil}
\affiliation{National Institute of Science and Technology for Complex Systems, 22290-180, Rio de Janeiro, Brazil}
\affiliation{Ibitipoca Institute of Physics - IbitiPhys, Concei\c{c}\~ao do Ibitipoca, 36140-000, MG, Brazil}

\date{\today}

\begin{abstract}
Rumors have ignited revolutions, undermined the trust in political parties, or threatened the stability of human societies. Such destructive potential has been significantly enhanced by the development of on-line social networks. Several theoretical and computational studies have been devoted to understanding the dynamics and to control rumor spreading. In the present work, a model of rumor-telling in opinion polarized networks was investigated through extensive computer simulations. The key mechanism is the coupling between ones' opinions and their leaning to spread a given information, either by supporting or opposing its content. We report that a highly modular topology of polarized networks strongly impairs rumor spreading, but the couplings between agent's opinions and their spreading/stifling rates can either further inhibit or, conversely, foster information propagation, depending on the nature of those couplings. In particular, a controversy-seeking mechanism, in which agents are stimulated to postpone their transitions to the stiffer state upon interactions with other  agents of confronting opinions, enhances the rumor spreading. Therefore such a mechanism is capable of overcoming  the  propagation bottlenecks imposed by loosely connected modular structures. 
 
\end{abstract}

\maketitle

\section{Introduction}
At the middle of $1789$, from July $20$ to August $6$, a rumor enigmatically spread like wildfire throughout France. The news was that outlaw bands were sweeping the prairies to cut the unripe wheat and destroy the crops. Rapidly, a massive panic wave -- the \textit{Grande Peur} (Great Fear) -- raised, transforming a rural commotion into an irreversible revolution~\cite{Hobsbawn} in France. Peasants plundered and set fire to landlords' properties, invaded registry offices to burn property deeds, pillaged churches and villages. Riots, attacks and fires simultaneously erupted in many provincial towns (e.g., Marseille, Lyon, Grenoble, Rennes, Le Havre and Dijon). Three weeks after the fall of the Bastille, French feudal social structure and its royal state machinery completely collapsed.

These iconic events strongly highlight the centrality of rumor spreading in human societies that were, to a lower or higher degree, ever self-organized as informational networks~\cite{Castells}. They also provide evidence that rumors can become or strategically be used to harm social stability. Hence, understanding the mechanisms and design means to regulate information dissemination are imperative tasks for social sciences and even economics~\cite{Kimmel,Schindler}.

For the $21$th century physics, in great measure focused on the emergence and propagation of information in out-of-equilibrium complex systems~\cite{Thurner2018}, the theoretical analysis of contact processes, epidemic spreading, and rumor dissemination became central to understand phase transitions, stochastic dynamics and irreversibility~\cite{Castellano2009,Pastor-Satorras}. Regarding the dissemination of information, in $1964$, inspired by the susceptible-infected-recovered (SIR) epidemic dynamics~\cite{anderson1992infectious} for the spreading of an infectious disease, Daley and Kendall reinterpreted and extended this epidemic model aiming to describe rumor-telling~\cite{Daley}. This pioneer model was extended in many directions by either adding traits to the original mechanism of rumor propagation or varying the structural properties of the underlying social networks~\cite{Castellano2009}. Thus, several classes (e.g., asymptomatic, debunkers, exposed, hibernators, and latent or skeptical) widen the classical ones -- spreader, ignorant, stifler -- present in the SIR model, leading to diverse rumor spreading models~\cite{Xia,Yang,Wang}. These models were designed to take into account hesitation, forgetfulness, trust, refutation, forced silence, education, and other human factors involved in realistic rumor propagation processes. Furthermore, the traditional approaches based on ordinary (spatially implicit) and partial (spatially explicit) differential equation models~\cite{Xia,Yang,Wang,Zhu} were joined to lattice and graph (homogeneous and heterogeneous) models~\cite{Zanette,Moreno,DeArruda2020}, characterized by exponential and power-law degree distributions, respectively~\cite{Pastor-Satorras}. These studies revealed that topological properties of a complex network substantially impact the dynamics of rumor propagation, particularly the reach and speed of information spreading.

Online communications networks neatly exhibit homophily which leads to a natural polarization in groups sharing distinct perspectives~\cite{Conover}. The mutual interactions within these groups create echo chambers, in which beliefs are reinforced due to repeated interactions with individuals sharing the same points of view, as observed, for instance, in the impeachment of the Brazilian president Dilma Roussef in $2016$~\cite{Cota2018} and French elections of $2017$ ~\cite{Gaumont}. Moreover, these communities form a new topological structure of the communication network as interconnected modules. So, superimposed to the heterogeneous degree distribution, there is an additional level of heterogeneity associated with community sizes in modular networks. Hence, the goal of the present paper is to investigate rumor spreading onto networks generated by adaptive opinion formation processes that lead to loosely connected modular networks forming echo chambers. A new content is released within a community of the polarized network and follows a rumor spreading process~\cite{Vega-Oliveros} coupled with the opinion of the interacting individuals according to different rules raging from beliefs' alignment to controversy-seeking where contrasting opinions hampers lost of interest on an issue. In the current paper, we show that, as would be expected, the highly modular structure of opinion polarized networks strongly impairs rumor spreading. However, the introduction of couplings between agent's opinions and their spreading/stifling rates has a striking effect on rumor-telling. Indeed, depending on the nature of those couplings, information propagation can be either further inhibited or enhanced up to the level observed in unpolarized networks, thus suppressing the modularity bottleneck.

The rest of the paper is organized as follows. In Section~\ref{sec:model} the model is presented. Initially, it is described how the polarized opinion networks considered in our analysis are generated. The mechanism for rumor spreading onto such networks is proposed. The simulation results obtained are reported in Section~\ref{sec:results} and discussed in Section~\ref{sec:discussion}. Finally, our major findings are summarized as well as some perspectives addressed in Section~\ref{sec:conclusions}. Additional methodological details are presented at~\ref{sec:methods}.  

\section{Model}
\label{sec:model}

The fundamental issue addressed in the present work is how a rumor spreads in a polarized opinion's networks formed previously in a possibly different context. For example, how an anti-vaccine rumor spreads onto an online network formed during political or ideological debates. As a concrete contemporary example, consider the on-line network formed by supporters or opponents of Brazilian president Jair Bolsonaro, who deliberately preached against vaccine safety, especially for kids. So, how strongly an anti-vaccine rumor, started within the bubble of supporters, would reach the opponents depending on how the individuals react to this content?

\subsection{Creating Polarized Networks}

The major trait of political polarization is the splitting of a social group in diverse subgroups sharing similar opinions and committed to a common ideology. Such communities self-organize in densely connected modules loosely interconnected among them. In order to generate networks with polarized opinions and modular topology, we used an adaptive network model with bounded confidence opinions, inspired in the classical Deffuant model~\cite{Deffuant}, proposed in Ref.~\cite{Maia}. A quick rundown of the model follows:

\begin{itemize}
	\item Initially, $N$ individuals supporting uniformly distributed opinions $o_i \in [0,1]$ are organized in a random network described by a power-law degree distribution $P(k)\sim k^{-\gamma}$. This network is generated according the uncorrelated-configuration-model (UCM)~\cite{Catanzaro}. An upper cutoff of $k_\text{max} = \sqrt{N}$ and a degree exponent of $\gamma = 2.7$ were used.
		
	\item In each time step, each individual computes its neighbor's average opinion $\langle o \rangle_i= \frac{1}{k_i(t)} \sum_{j \in \nu_i(t)} o_j(t)$ and how much it differs from his or her own opinion, $\Delta_i=o_i-\langle o \rangle_i$. If the difference is below the threshold $\epsilon_i$, he or she changes their opinions in order to resemble those of their neighbors. The opinion of each agent is updated as:
	\[ 
	o_i(t+1)=\left\{
	\begin{array}{lll}
	o_i(t)+\mu \Delta_i(t) & \mbox{if }  &\Delta_i(t) \leq \epsilon_i \\
	o_i(t) & \mbox{if } & \Delta_i(t) > \epsilon_i,\\
	\end{array}
	\right.
	\]
	where  $\mu$ is the opinion's convergence rate. The parameter $\epsilon_i$ is the tolerance threshold of agent $i$ to different opinions and is constant along the time. Low $\epsilon_i$ means low tolerance and increasing polarization.
	\item Every agent reconsiders their connections after opinion update. If any pair of connected individuals $i$ and $j$ sustains opinions differing by $\Delta_{ij}=|o_i-o_j|> \mbox{min}\{\epsilon_i,\epsilon_j\}$, they may break their connection with probability $p=1-e^{-\kappa\Delta_{ij}}$. 
	\item Broken connections can be reconnected at a future time step with probability $q=e^{-d_{ik}/d_0}$, if $\Delta_{ik}<\mbox{min}\{\epsilon_i,\epsilon_k\}$. This probability is a function of the distance $d_{ik}$ between the individuals $i$ and $k$. Here, $\kappa$ is the link rupture tendency, and  $d_0$ is the characteristic rewiring distance, assumed to be uniform for all agents.
\end{itemize}

\begin{figure}[hbt]
	\begin{center}
		\includegraphics[width=0.95\linewidth]{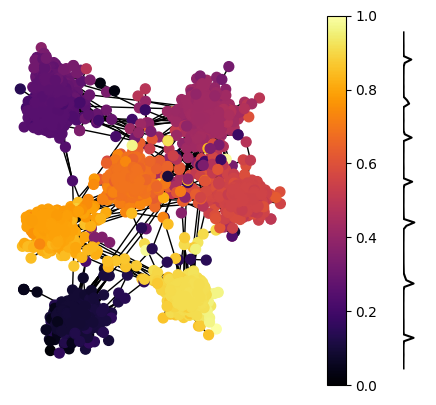}
	\end{center}
	\caption{Typical network obtained with the opinion formation model of Ref·~\cite{Maia}. Colors represent individual's opinions. The opinion distribution $p(o)$ is presented besides the color bar; the latter indicate the opinion scale in range $[0,1]$. The final number of individuals is around $10000$ and a special care was taken to pick only networks that maintain a balanced opinion distribution, i. e., $\langle o_i \rangle=0.5$ that does not favor any side. }
	\label{Net}
\end{figure}

Here, we assume $5\%$ of the individuals has a high tolerance of $\epsilon = 0.5$ while the remaining $95\%$ has a lower tolerance $\epsilon = 0.04$, distributed at random. We fixed $\mu = 0.8$, $\kappa = 3.5$, $d_0 = 4$ and an iteration time of $T = 120$.
This parameter set produces networks where a small number of individuals act as ``bridges'' between well connected communities of different opinions. Also, a balanced opinion distribution, i. e., $\langle o_i \rangle=0.5$ that does not favor any side is obtained. The initial size of $N = 18000$ produces a largest connected component of sizes around $N=10000$ individuals, in which the rumor spreading is investigated. Figure~\ref{Net} shows a typical  modular network with fixed opinions generated according to this protocol. Indeed, although in the original adaptive model~\cite{Maia} both opinions and connections change in time, here we use static networks with fixed opinions assuming that rumor spreading proceeds at a time scale much smaller than opinion formation processes.

\subsection{Opinion Driven Rumor Modeling}
Usually, rumor-spreading models divide the social group into three compartments in analogy to SIR epidemic dynamics~\cite{Daley,Pastor-Satorras,Vega-Oliveros}: ignorants (presented by S  in analogy with the susceptible individuals) that are completely unaware of the spreading information/rumor; spreaders (represented by I in analogy with infectious individuals) that actively disseminate the information, and stiflers (represented by R in analogy with the removed individuals) that is aware of the information but are no longer interested in spreading it around. Here we focus on the version introduced by~\cite{Maki} in which, upon pairwise social interactions, individuals can change their compartments according three different events:
\[
\text{I} + \text{S} \xrightarrow{\lambda} 2\text{I}, \hspace{0.5cm}
\text{I} + \text{R} \xrightarrow{\alpha} 2\text{R}, \hspace{0.5cm}
\text{I} + \text{I} \xrightarrow{\alpha} \text{R} + \text{I}.
\]
Ignorant individuals become aware with rate $\lambda$ when interacting with a spreader, while spreaders lose interest in the information with rate $\alpha$ when interacting with agents aware of the rumor. Differently from the SIR epidemic model, individuals do not recover spontaneously, but only if they perceive that spreading rumor is no longer novel nor interesting.

In realistic contexts, the members of a social group are very heterogeneous not only in their connectivity but also in their behaviors. So, individual's opinion is an important factor since one can choose either to spread the information or keep it depending on what he or she thinks about the rumor. For this reason, we altered the traditional rumor-spreading model in order to replicate such phenomena. The rate in which spreaders disseminate the rumor is now dependent on their alignment/opinion $o_i$ through a coupling function $f(o_i)$ . Specifically, each social agent has his own spreading rate per contact given by
\begin{equation}
\lambda_{i} = \lambda^* f(o_i),
\end{equation}
in which the parameter $\lambda^*$ is the average spreading rate. In turn, the rate in which an individual will grow bored of a rumor depends on both his own opinion $o_i$ and that of his interacting neighbor $o_j$ through another coupling function $g(o_i,o_j)$. Specifically, each link in the network has their own stifling rate given by
\begin{equation}
\alpha_{ij} = \alpha^* g(o_i,o_j),
\end{equation}
where $\alpha^*$ is the average stifling rate. The exact forms of the coupling functions $f(o_i)$ and $g(o_i,o_j)$ will naturally depend on the rumor in question. So, we propose few distinct functional forms for the coupling between opinions and the spreading/stifling rates. The \textbf{decoupled case}, corresponding to $f(o)=1$ or $g(o)=1$ depending whether one considers spreading or stifling processes, respectively, is also investigated for sake of comparison. Three coupling functions are considered: linear and unimodal coupling for spreading rate and controversy-seeking (CS) for stifling rate.

\textbf{Linear coupling.} In this model, the impetus to further propagate a new rumor is directly correlated with the individual's opinion and given by
\begin{equation}
f(o_i) = 2\eta_{\lambda} (o_i - \frac{1}{2}) + 1,
\end{equation}
where $\eta_{\lambda}\in[-1,1]$. The case $\eta_{\lambda}=0$ corresponds to the decoupled case while  $\eta_{\lambda} > 0$ or $<0$ favors opinions closest to $1$ or $0$, respectively. For instance, a rumor about vaccines causing side-effects would spread much more efficiently in anti-vaccination groups ($o \sim +1$) than in pro-vaccination ones ($o \sim 0$) implying $\eta_{\lambda} > 0$. So, the linear coupling results in a better spreading rate when there is an ideological alignment between the individual's opinion and the rumor.

\textbf{Unimodal coupling.} The function $f(o_i)$ has a non-monotonic symmetric form centered at $o = 0.5$ given by
\begin{equation}
f(o_i) = 1 + 2\eta_{\lambda}\left(|2o_i - 1| - \frac{1}{2}\right).
\end{equation}
Again, $\eta_{\lambda} \in [-1.+1]$ and $\eta_{\lambda} = 0$ means no coupling at all. But now $\eta_{\lambda} = 1$ favors opinions in either extremes ($o=0$ or $o=1$), whereas $\eta_{\lambda} = -1$ favors moderated opinions ($o=0.5$). Particularly, in the present work we are interested $\eta_{\lambda} > 0$ for which a rumor will be more effectively spread by both extremes. Indeed, while one extreme side agrees with the rumor and promotes its spread, the other extreme spreads it out while criticizing. In contrast, individuals supporting  moderate opinions do not get so excited about the rumor. An example would be a politician involved in a scandal, leading both opposition and supporters to debate the issue, sustaining distinct viewpoints, but still disseminating the news.

\textbf{Controversy-seeking coupling.} Assuming that opinion controversy, or rather polarization, can further stimulate and extend the debate concerning a rumor among the agents, we investigated the coupling
\begin{equation}
g(o_i,o_j)=Ae^{-\eta_{\alpha}|o_i-o_j|}
\end{equation}
for the stifling rate. It express that an interaction between individuals that know the rumor will remain stimulating if their opinions are sufficiently different, hindering them to turn into stiflers. Here, $\eta_{\alpha}$ value is always positive and $A$ is a normalization prefactor to guarantee the condition $\langle{g(o_i,o_j)}\rangle=1$. Consequently, the higher the coupling parameter $\eta_{\alpha}$, the more controversial the subject. Also, $\eta_{\alpha} = 0$ and $A=1$ correspond to opinion decoupled relations. Hereafter, when controversy-seeking coupling is used, we set $\eta_{\alpha} = 20$ and $A = 1.05$.

For each of these three types of coupling, the average values of spreading/stifling rates given by
\begin{equation}
\langle \lambda_i \rangle = \frac{1}{N} \sum_{i} \lambda_i = \lambda^*
\end{equation}
and
\begin{equation}
\langle \alpha_{ij} \rangle = \frac{1}{2N\langle k \rangle} \sum_{i\in \mathcal{V}(j)} \alpha_{ij} = \alpha^*.
\end{equation}
were kept the same independently of $\eta_{\lambda}$ and $\eta_{\alpha}$ through a suitable choice of $A$. 

\section{Results}
\label{sec:results}
\begin{figure}[hbt]
	\begin{center}
		\includegraphics[width=0.49\linewidth]{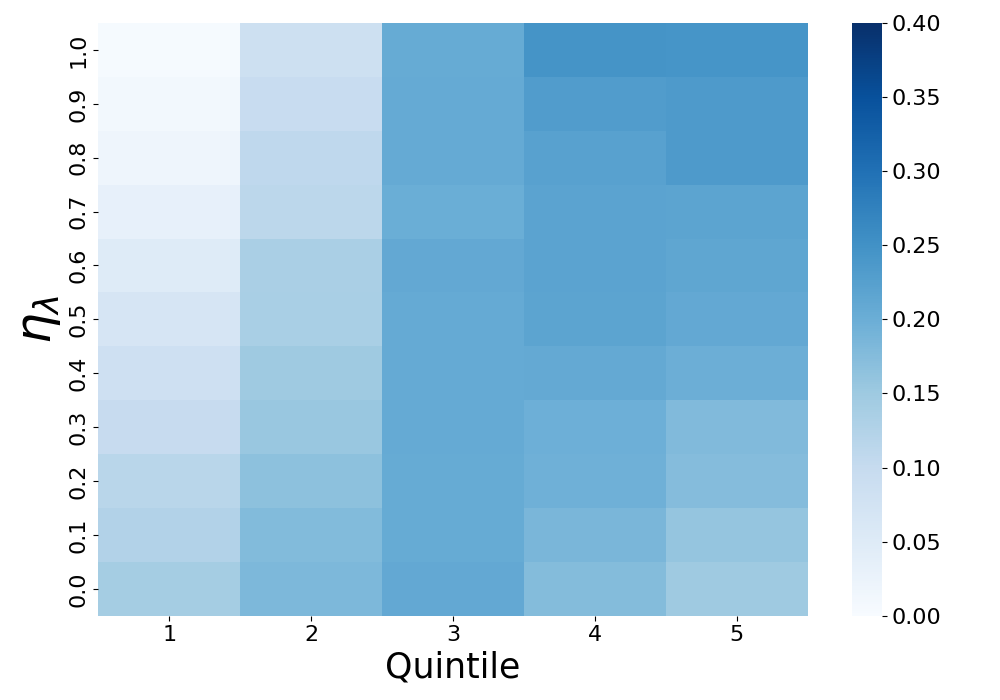}\includegraphics[width=0.49\linewidth]{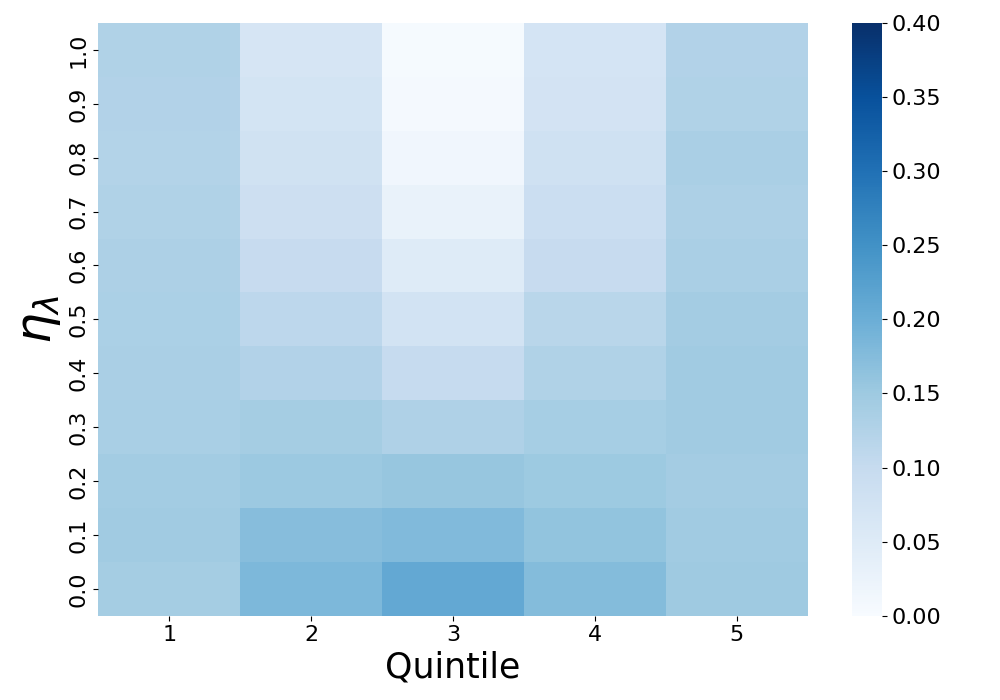}\\(a) \hspace*{4cm} (b)\\
		\includegraphics[width=0.49\linewidth]{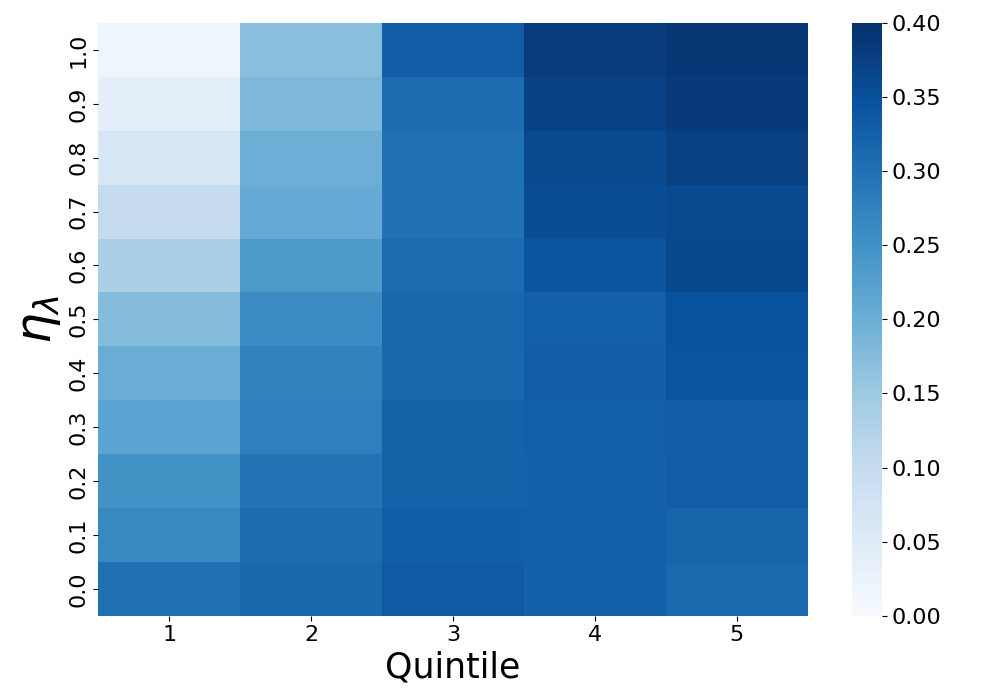}		\includegraphics[width=0.49\linewidth]{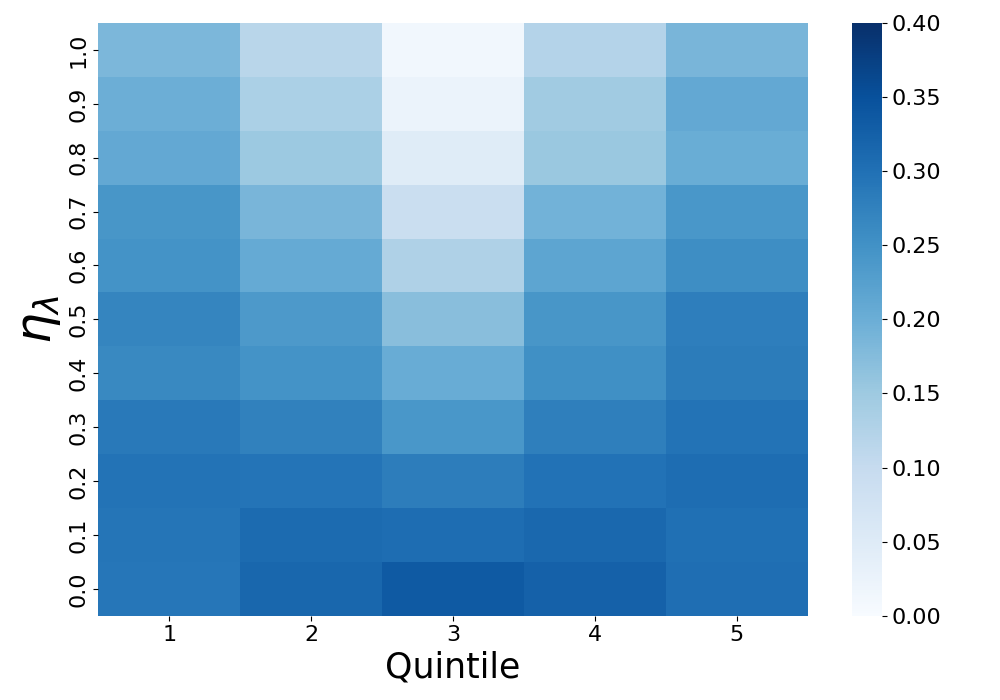}\\(c)\hspace*{4cm}(d)
	\end{center}
	\caption{Heat maps for final stifler density in different quintiles as function of the spreading coupling parameter $\eta_\lambda$. Color bars represent the density of stiflers. Different coupling models are shown: (a,c) linear  and (b,d) unimodal couplings (a,b) without and (c,d) with controversy-seeking.}
	\label{Eta_a}
\end{figure}
Simulations of rumor spreading were performed starting from a single spreader randomly chosen among the $N$ individuals comprising the social network.  All remaining agents are initially ignorants. The averages were computed for an ensemble of 50 independently generated networks and $400$ realizations of the dynamical process for each network.  For sake of reference, the spreading dynamics without  couplings between agents' opinions and their spread/stifling rates were considered. Furthermore, in order to address eventual effects on the rumor spreading dynamics induced primarily by the topological properties of polarized networks, numerical simulations were done on graphs after random rewiring preserving the degree of each node~\cite{barabasi2016network}. 

Since the opinion supported by every agent modulates its engagement in rumor spreading, we divided the entire opinion range into quintiles: $q=1$ ($0\leq o <0.2$), $\ldots$, $q=5$ ($0.8\leq o \leq 1$). The opinion distribution is almost uniform across the different quintiles with approximately 23.4\% of the nodes belonging to $q_3$, 19.6\% to $q_2$ and $q_4$, and 18.7 to $q_1$ and $q_5$.  The basic quantities used to characterize the rumor spreading were the fraction of stiflers $r_q$ of the $q$th opinion  quintile at the end of dissemination process. 

We investigated to what extent the coupling of opinion's agents with their spreading and stifling rates influences rumor-telling. Figures \ref{Eta_a}(a) and (b) show the fractions $r_q$ within the quintiles for dynamics with either linearly and unimodally coupled spreading rates and decoupled stifling rates. {In the absence of any coupling to agent opinion, a symmetric $r_q$ is expected with respect to the quintile centered on $o=1/2$, since the moderate opinion agents are the bridges between extremer opinion's agents and are reached by rumors starting into both sides.}  As one can observe in Figures~\ref{Eta_a}(a) and (b), positive $\eta_\lambda$ values increasingly enhance rumor spreading in quintiles supporting extreme opinions close to $1$, while impairing information dissemination in quintiles supporting opinions close to $0$. Evidently, the activation or inhibition of spreader activities is strengthened as the coupling constant $\eta_\lambda$ increases. In turn, the unimodal coupling facilitates the rumor spreading in the extreme quintiles, in which the opinions are near to $0$ and $1.0$. Moreover, as shown in Figures \ref{Eta_a}(c) and (d), the additional coupling of stifling rates with opinions strongly reinforces these tendencies. Therefore, each coupling model is actually doing the job it was designed for.

\begin{figure}[hbt]
	\centering
	\includegraphics[width=0.9\linewidth]{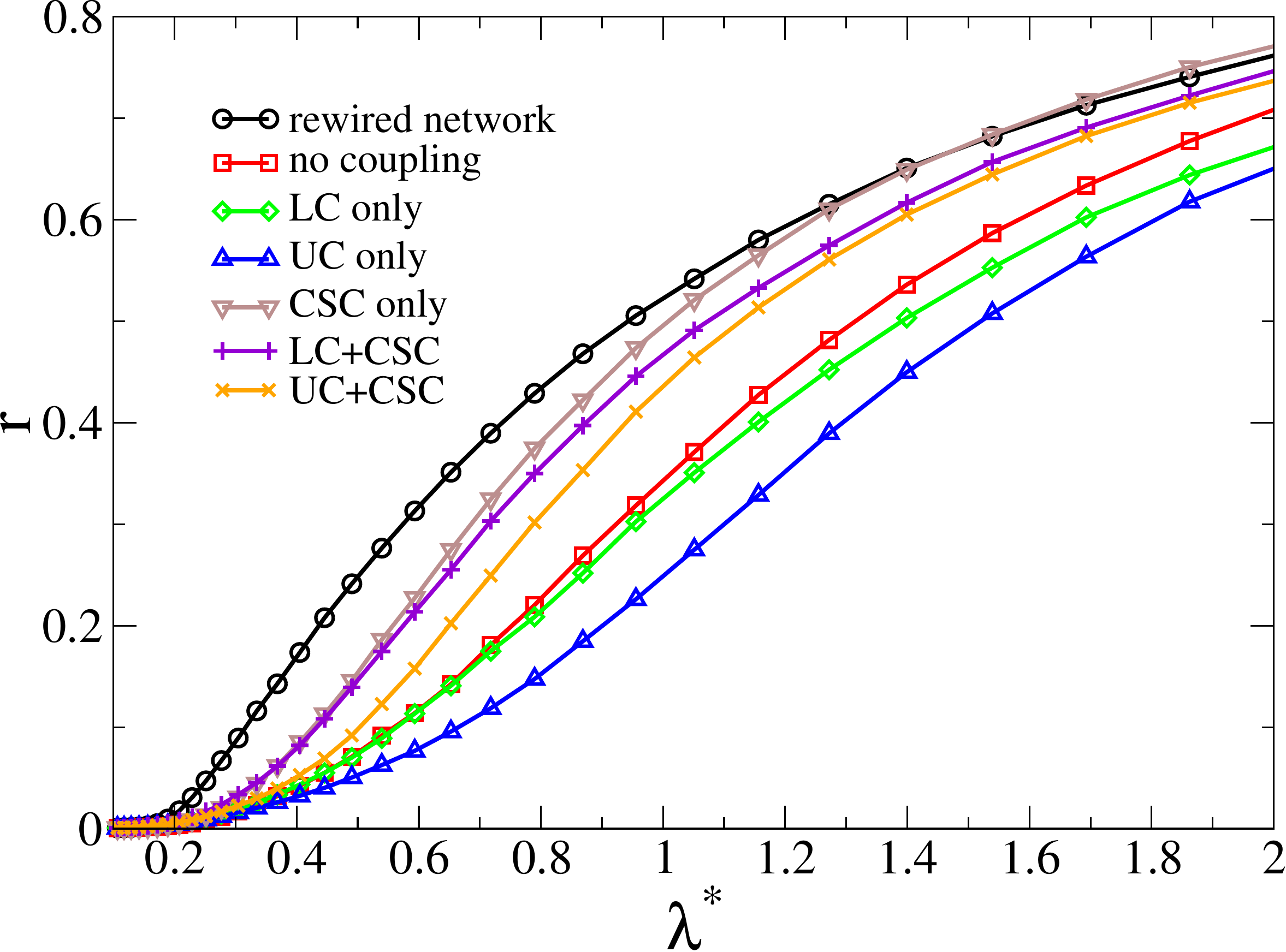}
	\caption{Final fraction of stiflers $r$ as a function of $\lambda^*$ for different types of opinion couplings. Networks size is $N=10^4$ individuals. The used parameter values are $\eta_{\lambda} = 0.5$, for linear and unimodal couplings, and $\eta_{\alpha} = 20$, $A=1.05$, for controversy-seeking coupling. Rewired networks are used as a control case, the opinion distribution is kept but links are reshuffled while preserving the degree distribution in order to eliminate the network modular structure. Acronyms: LC-linear coupling; UC- unimodal coupling; CSC - controversy-seeking coupling.}
	\label{fig:pR}
\end{figure}
The final global fraction of stiflers $r_\infty$ computed over the entire network  as function of the average spreading rate $\lambda^*$, for several coupling schemes,  is shown in Figure~\ref{fig:pR}. The first relevant outcome is seen from a comparison of rewired  (black curve) and modular (red curve) networks in the absence of coupling to opinions. The modular topology of opinion polarized networks impairs rumor dissemination in comparison to the randomized graphs. Whatever the coupling between agent opinion and its spreading rate ($\eta_{\lambda} > 0$), information dissemination is suppressed and the unimodal coupling (blue curve) has a greater impact than linear one (green curve). The controversy-seeking coupling significantly fosters rumor propagation,  regardless of the chosen spreading rate coupling, overcoming substantially the hurdles raised by network modular topology to information flow.

\begin{figure}[hbt]
	\begin{center}
		\includegraphics[width=0.95\linewidth]{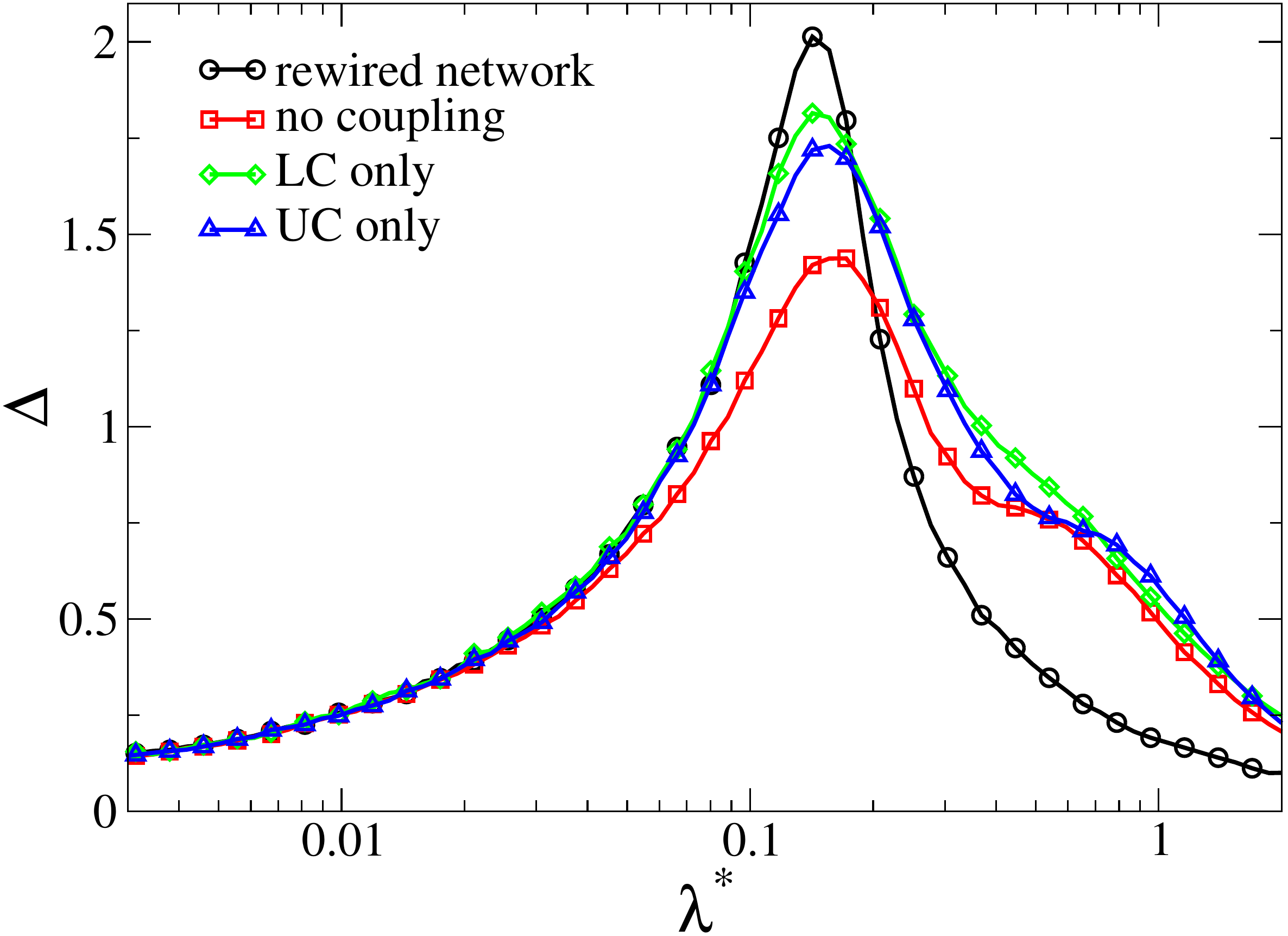}
		\includegraphics[width=0.95\linewidth]{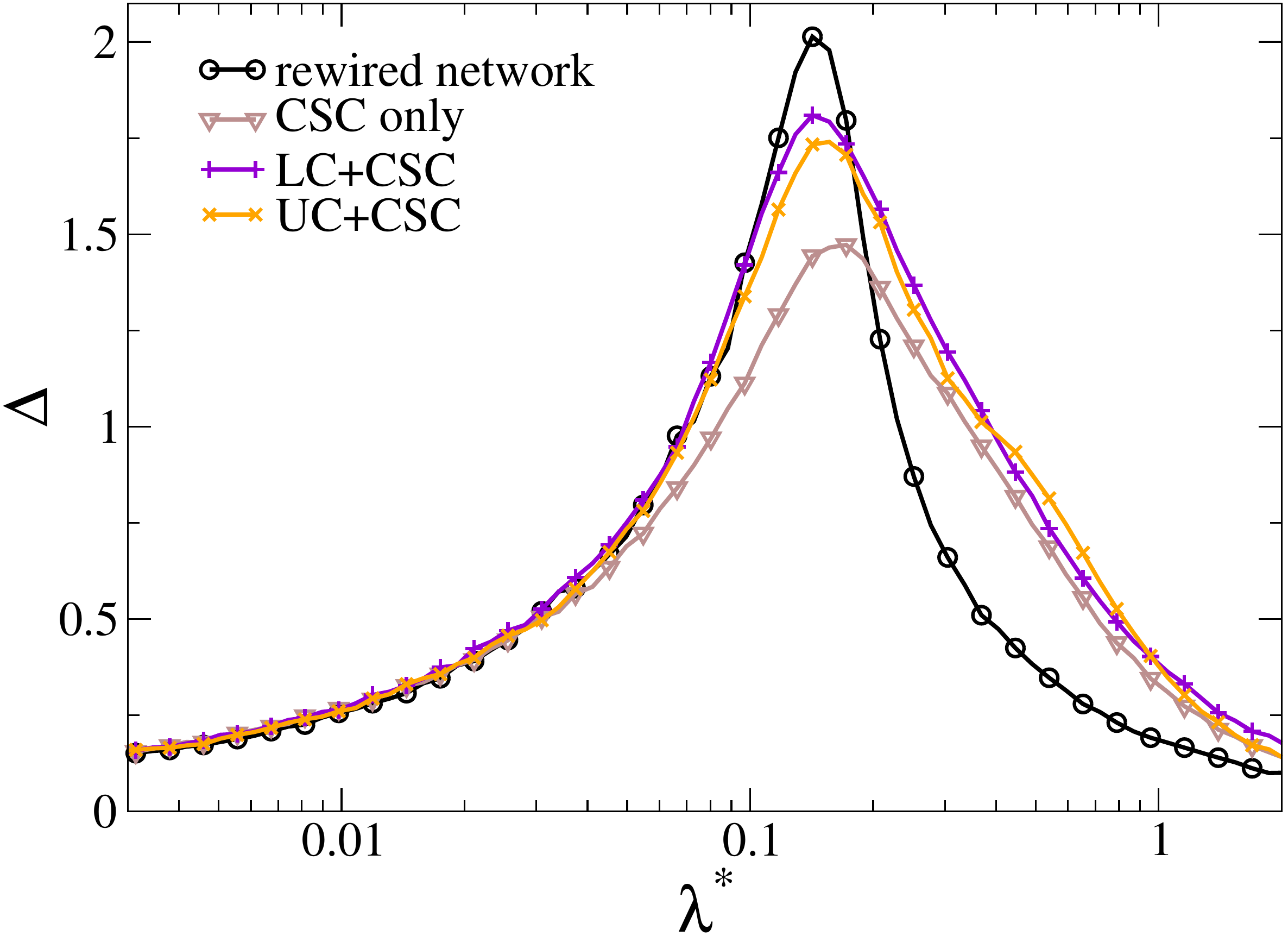}
	\end{center}
	\caption{Variability analysis for the  final density of stiflers for different rules of opinion coupling. Networks and parameters used are the same as in figure \ref{fig:pR}. Results are averaged over $10^4$ independent samples on $50$ different networks. Acronyms as in Fig.~\ref{fig:pR}.}
	\label{fig:Variab}
\end{figure}

The activation threshold is defined as a spreading rate $\lambda_\text{c}^*$ above which the rumor spreads to a finite fraction of network in the infinite size limit, while a localized outbreak of finite size (non-extensive) corresponding to vanishing fraction of the network is obtained for $\lambda^*<\lambda_\text{c}^*$. While dealing with finite-sizes in computer simulations, an effective threshold in SIR-like models can be be estimated using the peaks of the  variability as a function of the control parameter $\lambda^*$. The variability is defined as~\cite{Shu2015}
\begin{equation}
\Delta = \frac{\sqrt{\av{r^2_\infty}-\av{r_\infty}^2}}{\av{r_\infty}},
\end{equation}
where the averages are computed over the ensemble of independent initial conditions and polarized networks. The leading peak of variability curves presented  in Fig.~\ref{fig:Variab} are only slightly shifted  as different coupling rules are considered, being close to the the threshold of rewired (unstructured) network ($\lambda_\text{c}\approx 0.15$). This means that the initial outbreak activation depends essentially on  the local heterogeneity of the underlying network. A secondary effect is observed when only coupling of spreading rate with opinion is at work (absence of controversy seeking) manifested as a shoulder subsequent to the peak, enhanced for the unimodal and reduced for the linear coupling rules, in comparison with decoupled case. This result can be interpreted as the sudden activation of communities different from those where the rumor was initiated. The presence of controversy-seeking coupling rule promotes the inter-communicability among loosely connected communities and smears the secondary activation, as observed for epidemic spreading in  modular structures~\cite{Cota2018}.  

\begin{figure}[hbt]
	\begin{center}
		\includegraphics[width=0.95\linewidth]{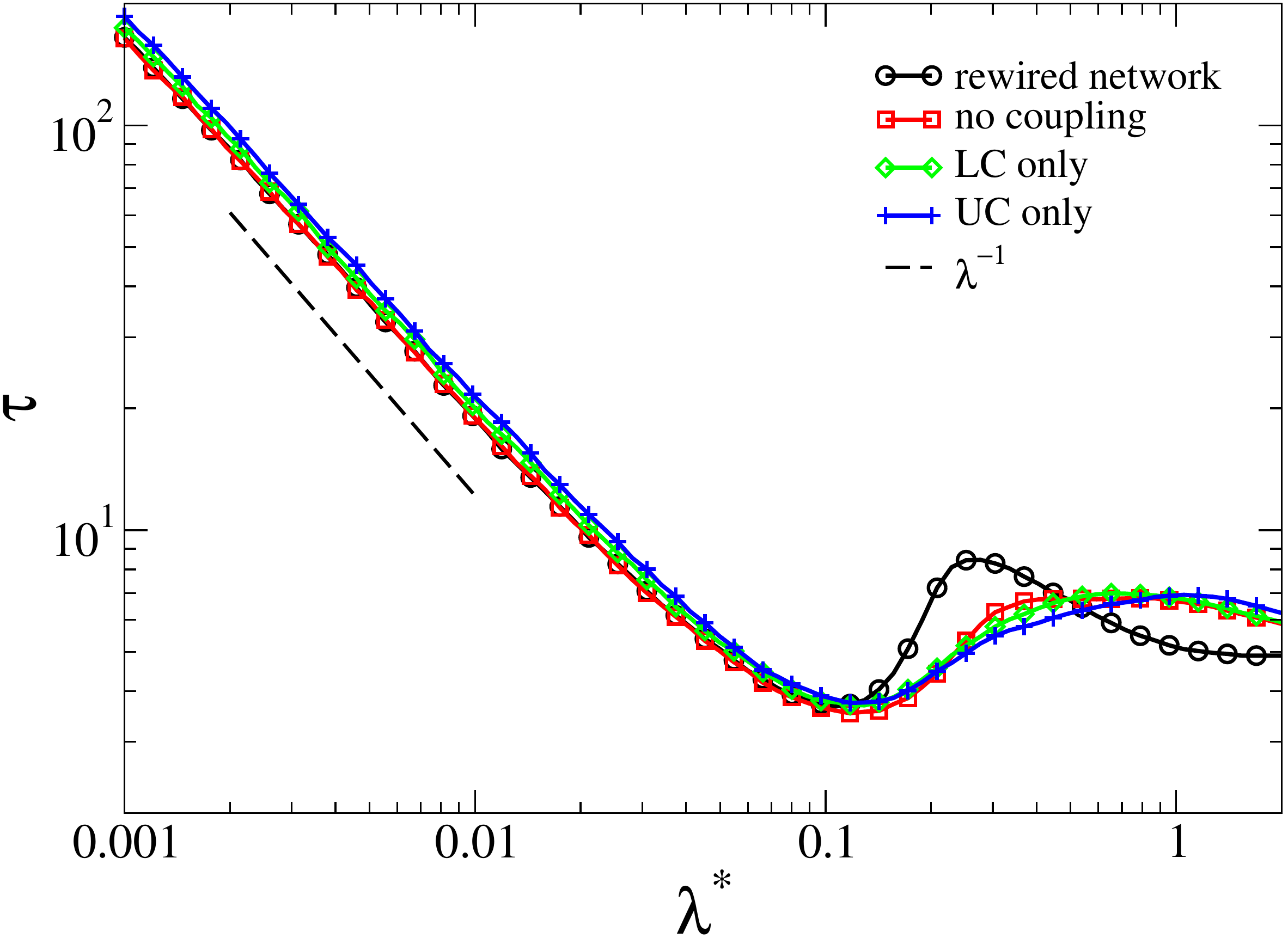}\\
		\includegraphics[width=0.95\linewidth]{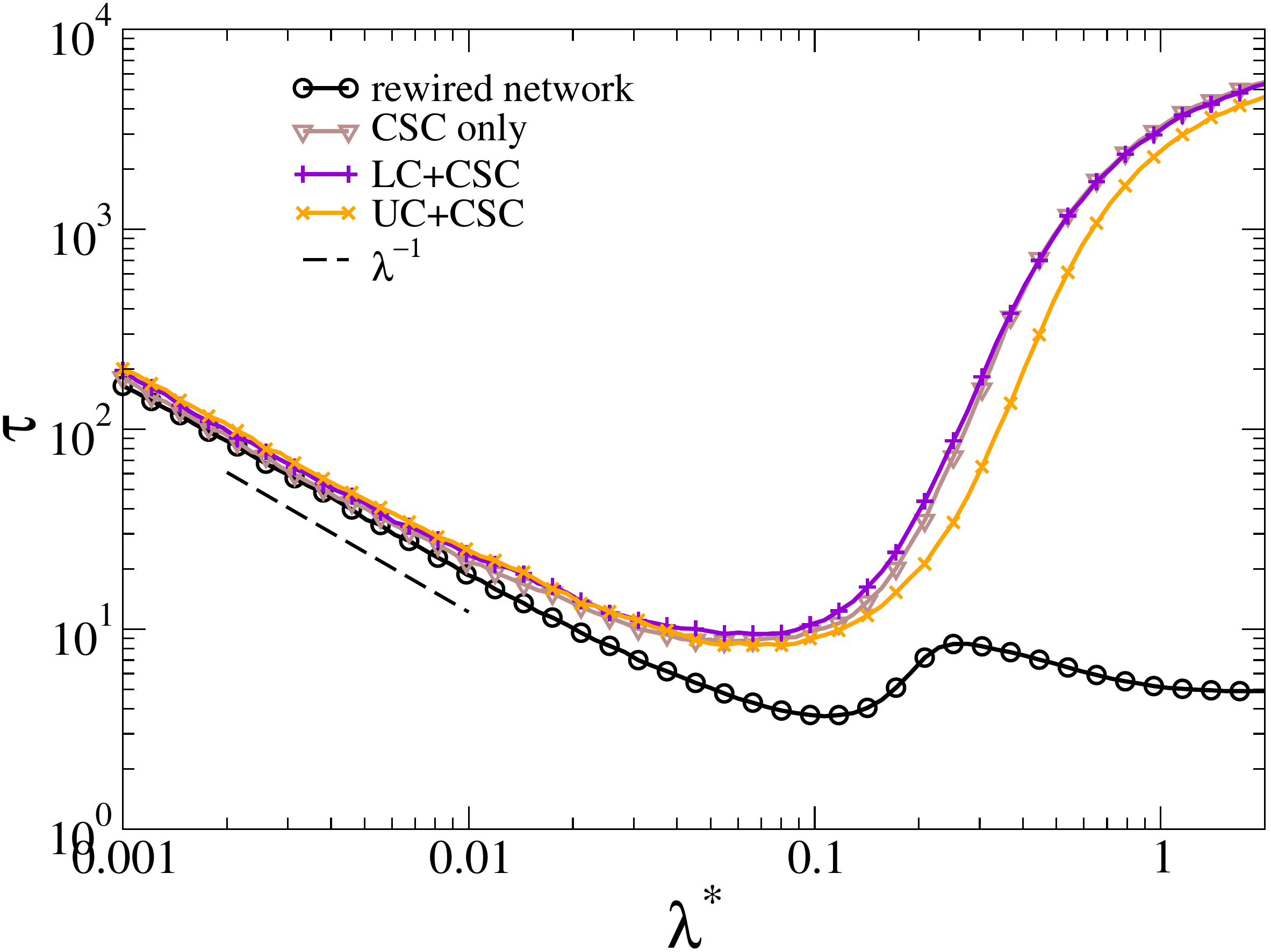}
	\end{center}
	\caption{Time to reach the absorbing state as a function of $\lambda^*$ and for different types of opinion couplings. Networks, parameter set, and acronyms are the same as in Fig.~\ref{fig:pR}. }
	\label{Tau}
\end{figure}

The total duration of the rumor spreading, defined as the elapsed time $\tau$ until the system reaches an absorbing state without spreaders, is presented in Fig.~\ref{Tau}. For small average spread rates, $\lambda^* \lesssim 0.2 \approx \lambda_\text{c}$, the expected scaling $\tau \sim \lambda^{*-1}$ holds (the greater $\lambda^*$, the faster the spreading)~\cite{FerrazdeArruda2022}, irrespective of the coupling rules, since the rumor remains essentially constrained to the network community in which it arose. For $\lambda^*>\lambda_\text{c}$, when the rumor is able to spread throughout distinct communities, the duration presents a local maxima qualitatively equivalent to the rewired network. This effect is hugely amplified by the controversy-seeking coupling, since agents supporting diagonally different opinions mutually stimulate each other to remain spreading the rumor state and delaying the global convergence into stiflers.  

\begin{figure*}[bth]
	\centering
	
\subfigure[\label{pb}]{\includegraphics[width=0.24\linewidth]{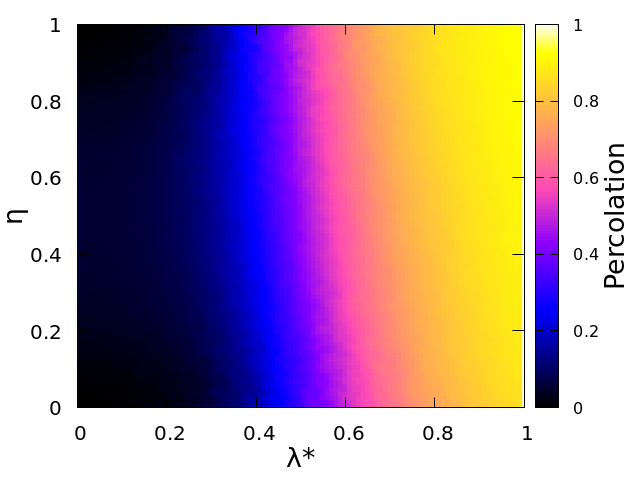}}
\subfigure[\label{pc}]{\includegraphics[width=0.24\linewidth]{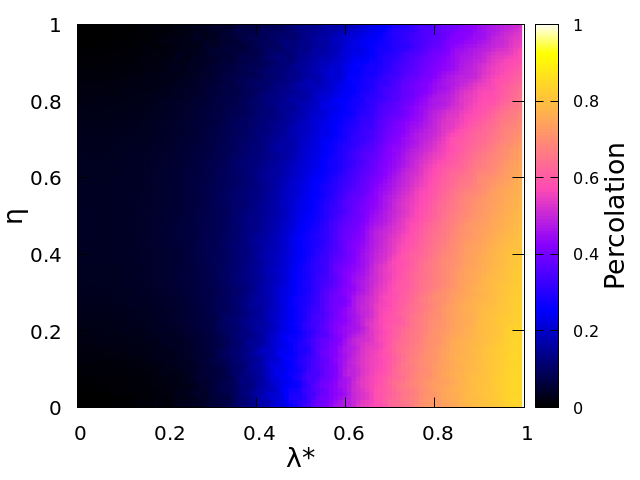}}
\subfigure[\label{pe}]{\includegraphics[width=0.24\linewidth]{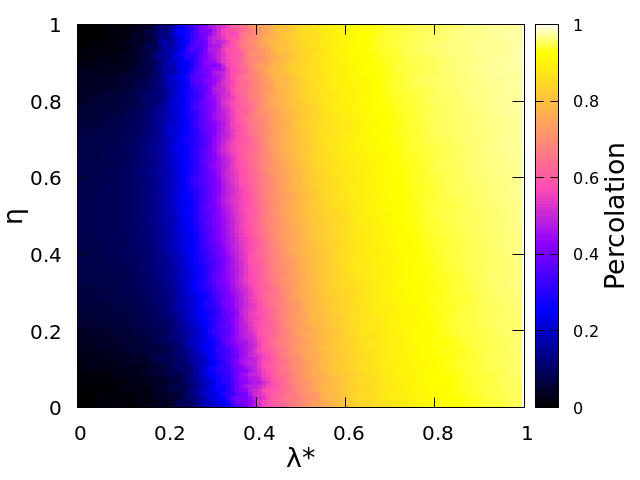}}
\subfigure[\label{pf}]{\includegraphics[width=0.24\linewidth]{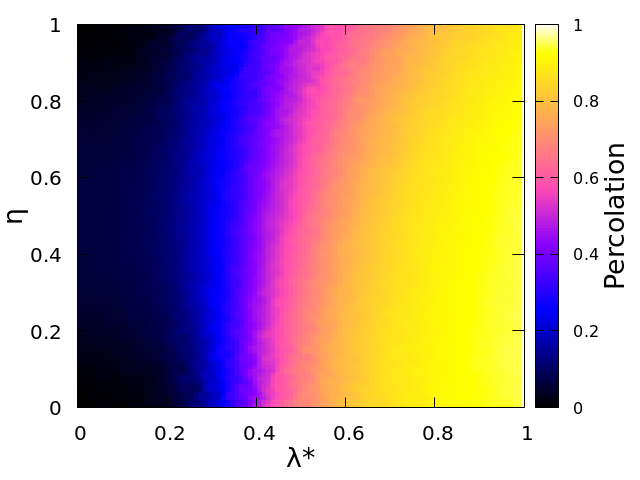}}\\
\subfigure[\label{qb}]{\includegraphics[width=0.24\linewidth]{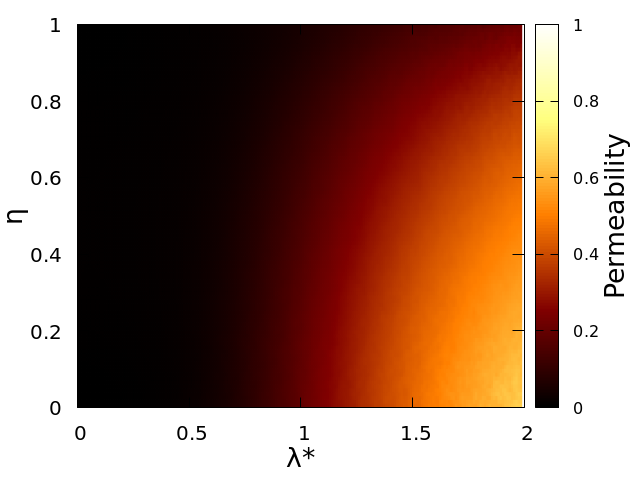}}
\subfigure[\label{qc}]{\includegraphics[width=0.24\linewidth]{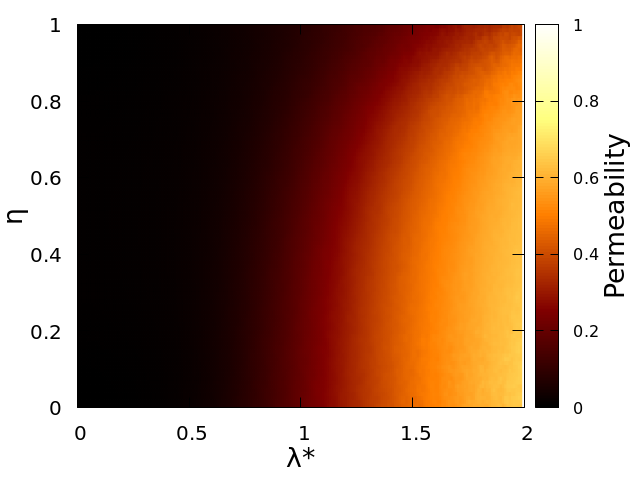}}
\subfigure[\label{qe}]{\includegraphics[width=0.24\linewidth]{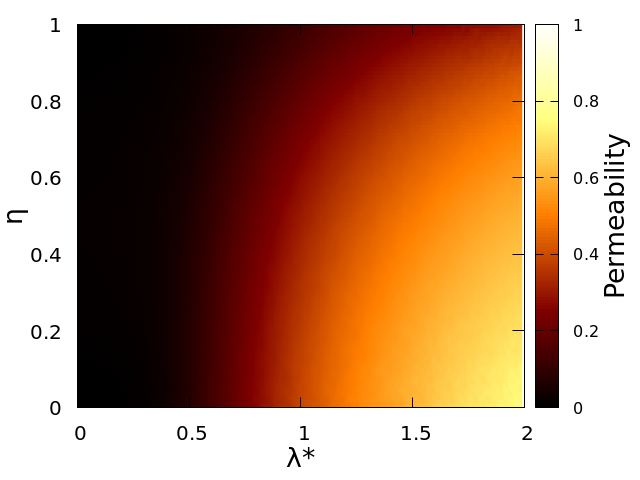}}
\subfigure[\label{qf}]{\includegraphics[width=0.24\linewidth]{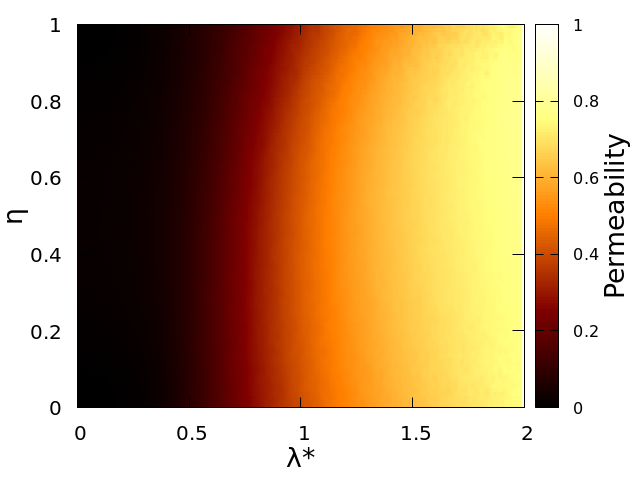}}
\caption{Heat maps in the space parameter $\lambda^*$ versus $\eta_{\lambda}$ for, given a rumor starting at the highest quintile ($o_i \in (0.8,1.0)$), (a-d) the probability that it will percolate, i.e, reach at least a single individual at the opposite quintile ($o_i \in (0.0,0.2)$) and (e-h) for the permeability $r_{1\rightarrow 5}$, defined as the fraction of individuals that are stiflers within the lowest opinion quintile. Four schemes of coupling to opinion are shown (acronyms as in Fig.~\ref{fig:pR}): (a,e) LC only, (b,f) UC only, (c,g) LC plus CSC, and (d,h) UC plus CSC.}
	\label{Qui}
\end{figure*}

We analyzed  the probability $p_\text{per}$ that at least one agent in the quintile $q_1$ gets aware of a rumor started in $q_5$. In other words, $p_\text{per}$ is a rumor percolation probability. We also investigate the permeability of rumors planted by distinct ideological shades. The permeability is hereafter defined as the average stationary fraction $r_{1\rightarrow5}$ of stiflers within the quintile $q_1$ ($o_i \in [0,0.2]$) generated by a rumor planted in the opposed extreme (quintile $q_5$, $o_i \in [0.8,1]$). Both percolation probability and permeability are shown in the parameter space $\eta_{\lambda}$ versus $\lambda^*$ for four coupling opinion models in Fig.~\ref{Qui}. One can see that permeability becomes appreciable  at spreading rates much higher than percolation probability does, happening for $\lambda^*\gg \lambda_\text{c}^*\approx 0.2$, the threshold above which the rumor is able to propagate at least within the community where it begun. This means, in general, that even when the rumor is able to escape from his original community, it is much harder to produce enough engagement outside its bubble.

\begin{figure}[hbt]
	\subfigure[\label{Pper_eta}]{\includegraphics[width=0.95\linewidth]{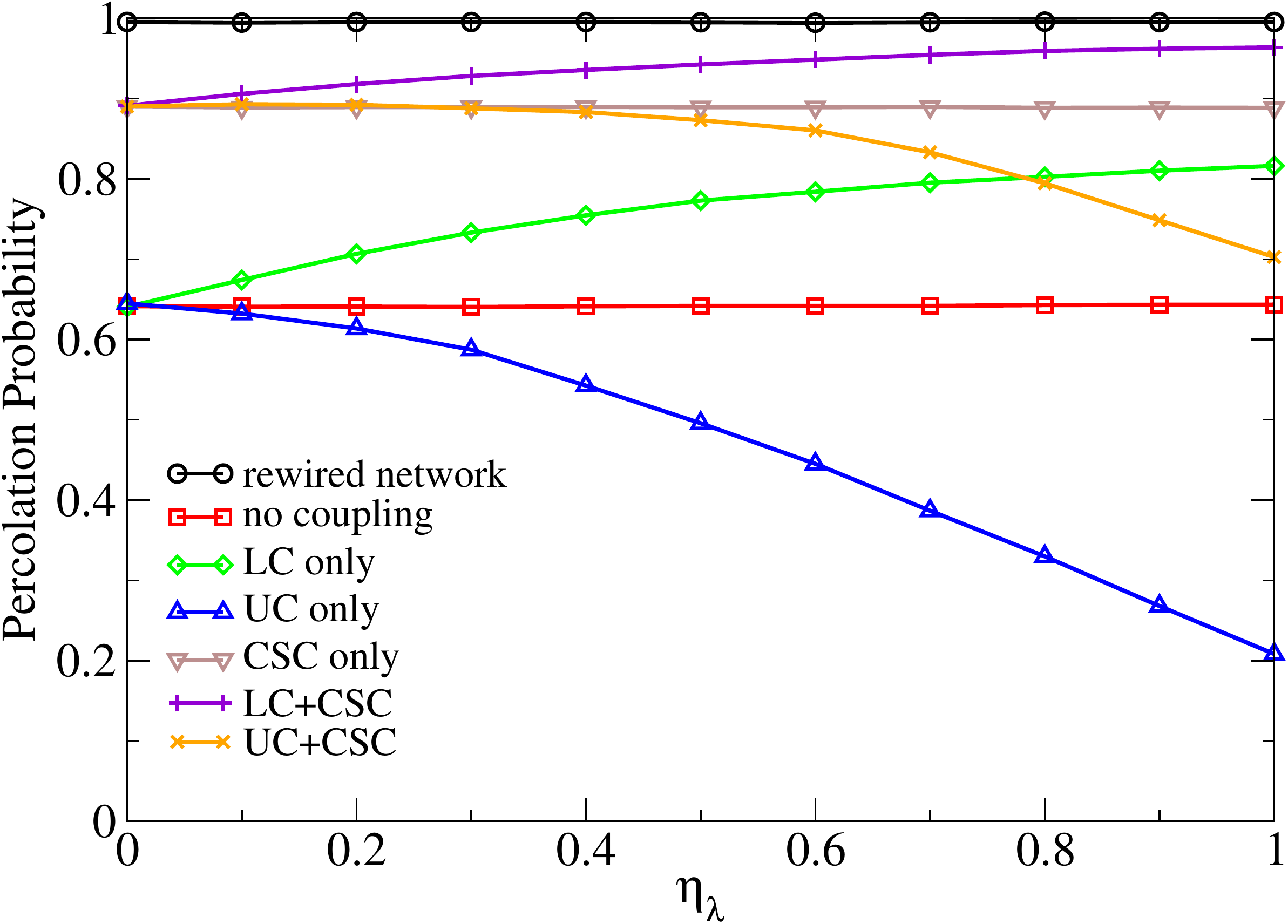}}
	\subfigure[\label{Perm_eta}]{\includegraphics[width=0.95\linewidth]{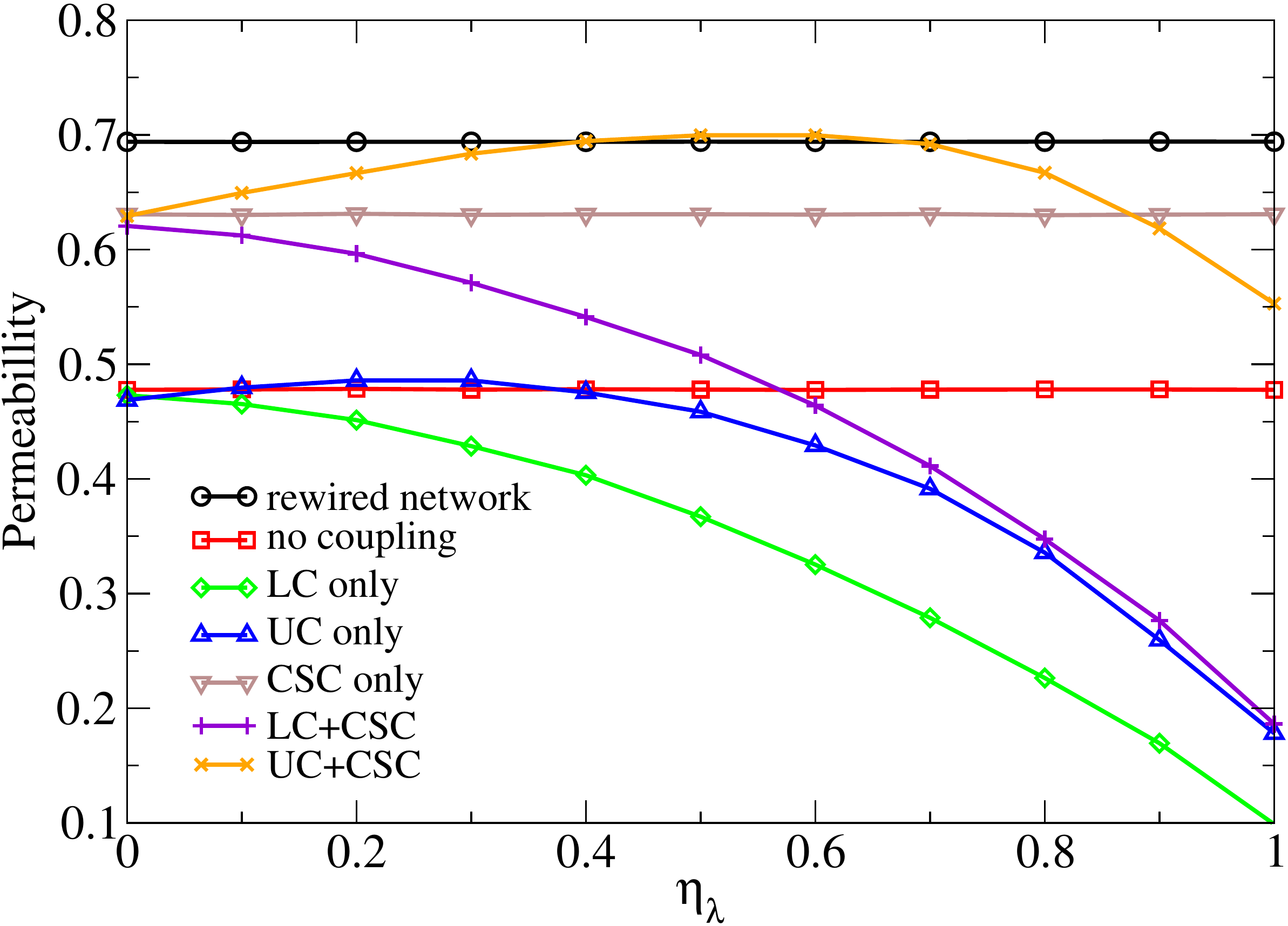}}
	\caption{(a) Percolation probability that a rumor reaches at least one individual of opposite opinion as a function of $\eta_{\lambda}$ for a fixed value of $\lambda^* = 0.6$. (b) Permeability $r_{1\rightarrow 5}$, defined as the fraction of individuals of extreme opinion that are reached for a rumor started in opposite opinion group, as a function of  $\eta_{\lambda}$ for fixed value for $\lambda^* = 1.5$.}
	\label{Eta}
\end{figure}

The role of different coupling rules is better visualized in cuts of fixed $\lambda^*$ shown in Figures~\ref{Pper_eta} and \ref{Perm_eta} showing percolation probability and permeability, respectively, as functions of the coupling parameter $\eta_{\lambda}$ for a fixed average spreading rate $\lambda^*$. The modular structure of the network, even without any coupling, drops percolation probability from 100\% to approximately 65\% in comparison with the rewired network. Linear coupling with spreading rate $\eta_\lambda$ enhances percolation chances, the more if concomitant with controversy-seeking. The unimodal coupling with spreading rate, however, drops the percolation probability if compared with controversy-seeking without coupling to spreading rate (down brown triangles) or no coupling at all (red squares). Figure~\ref{Perm_eta} shows that both linear and unimodal coupling impairs rumor permeability when acting without controversy-seeking if compared with the decoupled case (red squares). On the other hand, concomitant unimodal and controversy-seeking coupling rules enhances rumor permeability if compared with both controversy-seeking only (down brown triangles) and even being comparable with rewired network (black circles) without coupling where the bottlenecks for rumor were dropped. Note that linear-coupling presents apparently antagonistic results for percolation and permeability: while percolation is enhanced with more alignment with the original content the permeability is reduced for higher $\eta_{\lambda}$. However, the interpretation of this result is simple. While higher positive aligning helps to cross the borders of its community, the rumor does not lead to engagement outside its bubble.

\section{Discussion}
\label{sec:discussion}

Recently we investigated a model in which bounded confidence and homophily mechanisms drive both opinion dynamics and contact network evolution through either rewiring or breakage of social contacts~\cite{Maia}. In addition to the classical transition from global consensus to opinion polarization, this model exhibits a fragmentation cascade of the social network into echo chambers (communities) holding distinct opinions as the bridges interconnecting them are disrupted. Consensus is formed within each one of such echo chambers and the multiple surviving opinions are associated to distinct modules. This work, on one hand, and the dangerous mass events of rumor spreading witnessed worldwide in the last few years~\cite{Allcott,Grinberg2019,Marshall,Gaumont,Sokolova2018,Yerlikaya,Soares2019,Recuero_Soares_Gruzd_2020}, on the other hand, inspired us to address the issue of information propagation in societies strongly polarized by intolerance.

In the present work, we focused on a standard rumor model dynamics~\cite{Daley} with spreaders, stifflers and ignorants  onto loosely connected modular networks whose nodes represent the agents formed during an independent opinion formation process.  Spreading and stifling rates are coupled to the agent's opinions according to different rules that represent distinct reactions to a given content. The couplings for spreading rates consider either asymmetric (linear model) and symmetric (unimodal model) enhancement in relation to the individual opinion. The former rule represents an individual who propagates more content that he or she supports and less if he or she disagrees. The latter means that both extreme supporters and objectors to the content spread it more while moderate opinion individuals do it lesser.  To the best of our knowledge, these two ingredients, highly modular structures and transmission capacities coupled to agents' opinions, were not  considered concomitantly for modeling of rumor spreading.  Our results demonstrate that interplay between them produces great impacts upon rumor-telling.

Typical polarized opinion networks are structured in densely connected modules, loosely interconnected to each other by a reduced number of bridges, forming communities with high degree of homophily . Consequently, rumor spreading is strongly impaired in  modular structures since those bridges become stiflers breaking the intercommunication among communities. Moreover, this reduced contagion occurs even in the absence of any coupling between agent's opinion and spreading/stifling rates, but this purely structural effect can be enhanced or weakened depending on the coupling to opinions. Indeed, as shown in Figs.~\ref{fig:pR} and \ref{Tau},  coupling to opinion impairs while, in contrast, controversy-seeking coupling enhances rumor spreading onto polarized networks. 

Polarized networks are composed of dense modules loosely interconnected with each other by a small number of bridging agents having, on average, greater tolerance thresholds tending to sustain moderate opinions. Otherwise, they would not link individuals from distinct communities sharing different consensus. These bridges between communities are extremely important for rumor propagation, since in order to percolate through the communities, the rumor needs to cross bridging individuals. For unimodal (symmetric) couplings, where both extreme opinions lead to higher spreading rates while communities and bridging individuals of moderate opinions spread less than the average, it is difficult for the rumor to completely cross the network's modular structure to access the opposite extreme, thus lowering the percolation probability. But with linear (asymmetric) coupling, since rumors are created in communities ideologically aligned to it, they are able to easily acquire a large number of spreaders that are able to transmit the rumor to communities of moderate opinions, raising the percolation probability; see Figs.~\ref{Pper_eta} and \ref{Qui}.

The controversy-seeking coupling in stifling rates plays a fundamental role of keeping the rumor-telling active for very long times in a regime of high spreading rate, in contrast with the absence of coupling when individuals quickly get bored and become stiffler when exposed to the rumor several times. Because of their moderate opinion linking distinct communities, the controversy-seeking coupling maintains bridging individuals stimulated, consequently enhancing rumor spreading and permeability. So, the controversy-seeking coupling is in general very efficient in mitigating the handicaps of modular networks. The combination of unimodal (symmetric) and controversy-seeking coupling is especially efficient in rumor permeation since both extremes are interested in spreading the rumor while the bridges in between, though not as interested in spreading, still remain stimulated due to controversy; see Figs.~\ref{Perm_eta} and \ref{Qui}.

At last, we make some remarks concerning a qualitative ``validation'' of the present model by data from information spreading onto actual polarized social networks. Gaumont {\it et al}.~\cite{Gaumont} performed a detailed reconstruction of the French landscape during the $2017$ presidential elections from twitter data. Their results revealed a political environment highly fragmented and multi-polar represented by modular networks qualitatively very similar to those used here. A simple visual comparison between figure \ref{Net} here and figure $4$ in \cite{Gaumont} is enough. 
%
%
The structure of on-line communities constrained the information spreading among Twitters activists, and most retweet cascades occurred within one or two communities. This observation can, in principle, be solely explained by a highly modular network topology, as our simulational results evidence. But Gaumont {\it et al}. also observed that individuals supporting extreme political views are the most committed to their ideology. This finding provides empirical basis for our proposal of an unimodal coupling between the rumor spreading rate and the agent's opinion, spreading intensively both contents that are aligned with their ideology and contents that opposes their beliefs, the latter being spread as a form of criticism. According to our results for unimodal coupling, this ideological commitment further impair dissemination and circulation of ``polarized'' political information, particularly fake news as shown in blue curves of Fig.~\ref{Eta}. Therefore, unimodal coupling only is not sufficient to explain the twitter political debate during 2017 French elections. However, the introduction  of controversy seeking allows to overcome the bottlenecks imposed by moderate opinion agents leading to intense activity also in the ideologically opposed group.    Even not explicitly reported in Gaumont \text{et al.}~\cite{Gaumont}, our results for network permeability with controversy-seeking, orange curves in Fig.~\ref{Eta}, where a rumor traverses among extremist communities are qualitatively consistent with the observations in France  and support an important role played by  controversy seeking  in this debate.

\section{Conclusions}
\label{sec:conclusions}

Along history, rumors have ignited riots, flawed the trust in state institutions and political parties, shaken confidence in economic markets and corporations, as well as threatened the social stability of large human groups. Since the dawn of information societies, in which news propagate massively and rapidly through on-line social networks, understanding the dynamics and control of rumor-telling has become an imperative task. In the present paper, rumor spreading in opinion polarized social networks was investigated using computer simulations. In such networks the topology is highly modular with loosely connected communities and each of their communities are composed of nodes supporting very akin opinions. Furthermore, the rumor dynamics was extended in order to introduce spreading/stifling rates coupled to agent's (nodes) opinions.

Our simulations revealed that rumor spreading in opinion polarized networks is strongly impaired in comparison to rewired (unpolarized) networks. This is a purely topological effect since the communities present in polarized networks are interconnected by few bridges. So, due to the existence of a smaller number of paths connecting two randomly chosen nodes,  the information flow among modules ceases when the nodes bridging communities become stiflers. However, this scenario is significantly altered if agent's opinions and their spreading/stifling rates are coupled. Indeed, the unimodal (symmetric) coupling further inhibits rumor spreading because agents supporting moderate opinions have lower spreading rates. Consequently, rumors arising in one extreme opinion community spread to modules supporting moderate opinions, but tends to stop there and does not reach the opposed extreme. In contrast, the controversy-seeking coupling strongly enhances rumor spreading and, regardless of the chosen spreading rate coupling, can even overcome the hurdles raised by network modular topology to information flow. The reason is that nodes bridging modules, which connect agents supporting different opinions, are hindered to become stiflers under controversy-seeking coupling. Also, the interplay between these two behaviors --- unimodal and controversy-seeking --- can result in a continuous range of rumor-telling capacity in opinion polarized networks.

Finally, we speculate that unimodal and controversy-seeking couplings between agent's opinion and their spreading/stifling rates are the basic mechanisms underlying the major features observed in polarized political landscapes, such as that reconstructed from twitter data in French presidential elections in 2017~\cite{Gaumont}.

\appendix

\section{Methods}
\label{sec:methods}

The Monte-Carlo simulations are based on an optimized Gillespie algorithms with phantom processes~\cite{Cota2017}. Each edge and individual have different rates based on the heterogeneous opinion and degree distributions. The simulations are implemented as follows. A list with the labels of spreaders $\mathcal{V}^{(I)}$ is built and kept constantly updated along the simulations. We define the total attempt spreading rate of node $i$ given by $\mathcal{I}_i = k_i \lambda_{i}$ since $\lambda_{i}$ is the same for each of $i$'s links. The rate overestimates the actual spreading rate since only spreading  towards ignorants leads to effective transmission. The frustrated attempts to spreading to spreaders or stifflers constitute the the so-called phantom processes~\cite{Cota2017} which does not lead to changes of state but does contribute for time counting.  Similarly, the total stifling rate  of node $i$ is given by $\mathcal{R}_i = \sum_{j\in \nu_{i}} \alpha_{ij}$ with $\alpha_{ij} = \alpha^* g(o_i,o_j)$. The spreading events have a total rate given by $I = \sum_{i \in \mathcal{V}^{(I)}} \mathcal{I}_i$ while the two possible stifling events have total rate $R = \sum_{i \in \mathcal{V}^{(I)}} \mathcal{R}_i$ each.

The simulation is then implemented as follows. At each time step, a spreading event is selected with probability $I/(I + 2R)$. A spreader $i$ is chosen from the list $\mathcal{V}^{(I)}$ with probability proportional to $\mathcal{I}_i$ and  one of its neighbors $j$ is chosen at random with equal chance.  If $j$ is an ignorant then he or she becomes a spreader. If not, no change of state is implemented. Any of the two stifling possible events, $I + R \to 2R$ or $I + I \to I+R$, are selected with the same probability $R/(I + 2R)$. Similarly to the spreading events, a spreader and one of his or her neighbors are chosen with probabilities proportional to $\mathcal{R}_i$ and $\alpha_{ij}$, respectively.  If the event $I + R \to 2R$ was selected, $i$ becomes a stiffler only if $j$ is also a stifler, while if the event $I + I \to R + I$ was selected, $i$ only becomes a stifler if $j$ is a spreader. If these conditions are not met, no change of state is implemented. The time is incremented by $dt = -\ln(\xi)/(I + 2R)$ after every step whether states were altered or not,  where $\xi$ is a random number uniformly distributed in the range $(0,1)$.

\bigskip

\section*{Acknowledgments}
{This work was partially supported by the Brazilian agencies \textit{Conselho Nacional de Desenvolvimento Científico e Tecnológico}- CNPq (Grants no. 430768/2018-4 and 311183/2019-0) and \textit{Fundação de Amparo à Pesquisa do Estado de Minas Gerais} - FAPEMIG (Grant no. APQ-02393-18).} This
study was financed in part by the \textit{Coordenação de Aperfeiçoamento de 		Pessoal de Nível Superior} (CAPES) - Brasil  - Finance Code 001.

\bibliographystyle{elsarticle-num.bst}
\bibliography{bib}

\end{document}